\documentclass[aps,prb,reprint,amsmath,amssymb,superscriptaddress]{revtex4-2}
\usepackage{amsmath} 
\usepackage{graphicx}
\usepackage{dcolumn}
\usepackage{bm}
\usepackage{physics}
\usepackage{color}
\usepackage{booktabs}
\usepackage[colorlinks=true, citecolor=blue, linkcolor=blue, urlcolor=blue]{hyperref}
\usepackage{color}
\usepackage{graphicx}
\usepackage{algpseudocode}
\usepackage[compat=1.1.0]{tikz-feynman}

\usepackage{tikz}
\usepackage[compat=1.1.0]{tikz-feynman}

\begin{document}

\preprint{APS/123-QED}

\title{Combining Electron-Phonon and Dynamical Mean-Field Theory Calculations of Correlated Materials: Transport in the Correlated Metal Sr$_2$RuO$_4$} 

\author{David J. Abramovitch}%
\thanks{D.J.A and J.-J.Z. contributed equally to this work}
\affiliation{Department of Applied Physics and Materials Science, \protect\\ California Institute of Technology, Pasadena, California 91125}
\author{Jin-Jian Zhou}
\thanks{D.J.A and J.-J.Z. contributed equally to this work}
\affiliation{School of Physics, Beijing Institute of Technology, Beijing, 100081, China}
\author{Jernej Mravlje}
\affiliation{Jožef Stefan Institute, Jamova 39, 1000 Ljubljana, Slovenia}
\author{Antoine Georges}
\affiliation{Coll\`ege de France, Paris, France and CCQ-Flatiron Institute, New York, NY, USA}
\author{Marco Bernardi}
\email{bmarco@caltech.edu}
\affiliation{Department of Applied Physics and Materials Science, \protect\\ California Institute of Technology, Pasadena, California 91125}
\affiliation{Department of Physics, California Institute of Technology, Pasadena, California 91125}%

\begin{abstract}
Electron-electron ($e$-$e$) and electron-phonon ($e$-ph) interactions are challenging to describe in correlated materials, where their joint effects govern unconventional transport, phase transitions, and superconductivity. 
Here we combine first-principles $e$-ph calculations with dynamical mean field theory (DMFT) as a step toward a unified description of $e$-$e$ and $e$-ph interactions in \mbox{correlated materials.}
We compute the $e$-ph self-energy using the DMFT electron Green's function, and combine it with the $e$-$e$ self-energy from DMFT to obtain a Green's function including both interactions. This approach captures the renormalization of quasiparticle dispersion and spectral weight on equal footing. 
Using our method, we study the $e$-ph and $e$-$e$ contributions to the resistivity and spectral functions in the correlated metal Sr$_2$RuO$_4$. 
In this material, our results show 
that $e$-$e$ interactions dominate transport and spectral broadening in the temperature range we study (50$-$310~K), while $e$-ph interactions are relatively weak and account for only $\sim$10\% of the experimental \mbox{resistivity.}
%
We also compute effective scattering rates, and find that the $e$-$e$ interactions result in scattering several times greater than the Planckian value $k_BT$, whereas $e$-ph interactions are associated with scattering rates lower than $k_BT$.
Our work demonstrates a first-principles approach to combine electron dynamical correlations from DMFT with $e$-ph interactions in a consistent way, advancing quantitative studies of correlated materials.
\end{abstract}

\vspace{-20pt}
\maketitle
%
In strongly correlated materials, characteristic behaviors such as high-temperature superconductivity~\cite{pickett_electronic_1989,kordyuk_iron_2012}, phase transitions~\cite{quantum-phase, heavy-fermion}, multiferroicity~\cite{multiferroic}, and unconventional transport~\cite{deng_shining_2014,deng_how_2013} involve a subtle interplay of charge and lattice degrees of freedom. 
The atomic vibrations (phonons) couple with the strongly interacting $d$ or $f$ electrons, resulting in electron-electron ($e$-$e$) and electron-phonon ($e$-ph) interactions with nontrivial dependence on orbital, spin, and crystal symmetry~\cite{iwasawa_interplay_2010, van_loon_competing_2018}. 
Although heuristic models can address this rich phenomenology \cite{bardeen_electron_phonon_1955, sangiovanni_electron-phonon_2005,sangiovanni_electron-phonon_2008,kuchinskii_interplay_2009,cai_antiferromagnetism_2021,xing_qmc_2021},  deriving generic quantitative frameworks to combine $e$-$e$ and $e$-ph interactions remains challenging~\cite{Ziman,mahanManyParticle2000}, especially in correlated materials.
\\ 
\indent 
First-principles calculations based on density functional theory (DFT) and its linear-response extension, density functional perturbation theory (DFPT), can address the electronic structure~\cite{Martin-book}, lattice dynamics~\cite{baroni_phonons_2001} and $e$-ph interactions~\cite{bernardi2016first} in many materials. 
Yet these methods often fail to describe important features of correlated systems~\cite{cohen2008insights}, predicting incorrect ground states, lattice vibrations, and $e$-ph coupling~\cite{Cococcioni-dfptu,zhou_abinitio_2021}. 
\mbox{Recent} work has focused on two directions to improve the description of $e$-ph interactions in correlated materials:    
Hubbard-corrected DFT (DFT+U), which has enabled calculations of $e$-ph interactions in a Mott insulator~\cite{zhou_abinitio_2021}, and DFPT with improved electronic correlations (from GW or hybrid functionals) which has revealed correlation-enhanced $e$-ph interactions in metallic systems~\cite{yin_correlation-enhanced-2013,wen_unveiling_2018, li_electron_2019,li_unmasking_2021}. 
\\
\indent  
However, while GW and DFT+U can renormalize the band structure and $e$-ph coupling, both methods describe the electronic states in the quasiparticle (QP) picture, mapping them to noninteracting bands. 
These approximations are typically used as a starting point to study equilibrium properties and nonequilibrium dynamics~\cite{zhou_abinitio_2021,li_unmasking_2021}. 
Developing first-principles calculations of $e$-ph interactions beyond the QP picture remains challenging~\cite{bernardi2016first,giustino_electron_phonon_2017}. Ideally, one would use the full electron Green's function renormalized by $e$-$e$ interactions, which includes the renormalization of both QP weight and band dispersion, as a starting point to study $e$-ph interactions in correlated materials~\cite{petukhov_correlated_metals_2003,kent_toward_2018}. 
\\
\indent
Dynamical mean field theory (DMFT) and its \textit{ab initio} variant, DFT+DMFT, can capture dynamical electronic correlations by mapping the solid to an embedded atomic site with a self-energy local in the atomic orbital basis~\cite{georges_dynamical_1996,kotliar_electronic_2006,Kotliar-PhysToday}. 
These methods have been successful in describing the ground state and transport properties in materials with strongly interacting $d$ or $f$ electrons, including both correlated metals and insulators~\cite{Georges_strong_correlations_2013, nowadnick_quantifying_2015, xu_hidden_2013,deng_shining_2014, deng_transport_2016}. A key question addressed in this work is how one can combine first-principles DMFT and $e$-ph calculations to explicitly treat $e$-$e$ and $e$-ph interactions together in correlated materials.
\\
\indent 
Here we show first-principles calculations combining $e$-ph and DMFT $e$-$e$ interactions using a Green's function approach. We compare two treatments of the $e$-ph interactions $-$ the Fan-Migdal self-energy with standard approximations~\cite{bernardi2016first} (using DFT or DMFT-renormalized band structures) and the same self-energy  computed as a convolution integral of the DMFT Green's function. As a proof of principle, we apply this method to Sr$_2$RuO$_4$ (SRO) in the normal state, a prototypical correlated metal~\cite{mackenzie_quantum_1996,maeno1997two,tyler_high-temperature_1998,bergemann_detailed_2000,sakita_anisotropic_2001, iwasawa_interplay_2010, stricker_optical_2014,stadler_dmft_nrg_2015,deng_transport_2016, wang_quasiparticle_2017, kim_spin_orbit_2018,zingl_hall_2019,tamai_high_resolution_2019,linden_imaginary_time_mps_2020, kugler_strongly_2020, cao_tree_2021} (and unconventional superconductor below 1 K~\cite{mackenzie_superconductivity_2003,mackenzie_even_2017,beck_effects_2022}) 
requiring treatment of strong $e$-$e$ interactions between $d$ orbitals~\cite{tamai_high_resolution_2019,cao_tree_2021,linden_imaginary_time_mps_2020, kugler_strongly_2020}. 
%
%
In SRO, we find that the $e$-ph interactions are relatively weak and momentum dependent, in contrast with the strong local $e$-$e$ interactions. 
From Green-Kubo calculations, we find that $e$-$e$ interactions account for $\sim$50\%, and $e$-ph interactions $\sim$10\%, of the experimental \mbox{resistivity.} 
Only in the $e$-ph calculation using the DMFT Green's function the resistivity equals the sum of the $e$-ph and $e$-$e$ contributions; 
in contrast, the standard $e$-ph calculation with renormalized QP bands leads to an artificial enhancement of $e$-ph interactions and nonadditive resistivities. The origin of these trends is discussed in detail, together with possible improvements for resistivity calculations in correlated materials and future extensions of our method.\\

\section{Numerical Methods}
\vspace{-10pt}
\label{sec:dftdfptdmft}
We compute the electron Green's function by combining the DMFT and $e$-ph self-energies: 
\begin{equation}
   G_{n\mathbf{k}}(\omega,T) = [\omega + \mu(T) - \varepsilon_{n\mathbf{
   k}} - \Sigma_{n\mathbf{k}}^{\rm e-e}(\omega,T) - \Sigma_{n\mathbf{k}}^{\rm e-ph}(\omega,T)]^{-1}
   \label{eq:Gtotal}
\end{equation}
where $T$ is temperature, $\varepsilon_{n\mathbf{k}}$ are electronic band energies, $\omega$ is the electron energy, $n$ is a band index, $\mathbf{k}$ is crystal momentum, and $\mu$ is a temperature dependent chemical potential obtained from DMFT. 
This Green's function includes the DMFT $e$-$e$ self-energy in the band basis, $\Sigma_{n\mathbf{k}}^{\rm e-e}(\omega,T)$, which is computed starting from DFT, and the $e$-ph self-energy $\Sigma_{n\mathbf{k}}^{\rm e-ph}(\omega,T)$, which we compute using different approximations as discussed below.  
\\
\indent 
Our calculations in SRO use a DMFT self-energy taken from recent work~\cite{tamai_high_resolution_2019} and transformed from the Wannier-orbital to the band basis using~\cite{Marzari-RMP}
\begin{equation}
      \mathbf{\Sigma}_{\mathbf{k}}^{\rm e-e, b} = \mathbf{U}_\mathbf{k}^{\dagger}\, \mathbf{\Sigma}^{\rm e-e,w}\, \mathbf{U}_\mathbf{k} 
      \label{eq:sigmaband}
\end{equation}
where $\mathbf{U}_\mathbf{k}$ are unitary matrices made up by eigenvectors of the Wannier Hamiltonian, while $\mathbf{\Sigma}^{\rm e-e,w}$ and $\mathbf{\Sigma}_\mathbf{k}^{\rm e-e,b}$ are respectively the self-energies in the Wannier and band basis; the former is calculated directly from DMFT and is therefore \mbox{$\mathbf{k}$-independent.} As in the DMFT calculation, both the Wannier and band basis self-energies are taken to be diagonal~\footnote{This assumption is often justified $-$ in SRO, for example, the off-diagonal elements of the DMFT self-energy have been shown to be negligibly small in Ref.~\cite{tamai_high_resolution_2019}.}.
\\
\indent
These DMFT calculations use a 3-orbital correlated subspace with $t_{2g}$ symmetry resulting from the hybridization of Ru 4$d$ ($d_{xy}$, $d_{yz}$, $d_{xz}$) and O 2$p$ orbitals; the orbitals interact with Hund's type Coulomb repulsion and exchange~\cite{tamai_high_resolution_2019}. The DMFT impurity problem is solved on the imaginary-time axis with the TRIQS/CTHYB quantum Monte Carlo solver~\cite{gull_ctmc_2011,seth_triqscthyb_2016} and analytically continued to the real-frequency axis using Padé approximants~\cite{parcollet_triqs_2015}. These methods are discussed and validated in greater detail in recent work~\cite{tamai_high_resolution_2019}.  
Previous studies have clarified the role of spin-orbit coupling (SOC) in SRO~\cite{haverkort_strong_2008,veenstra_spin_orbital_2014, zhang_fermi_surface_2016, kim_spin_orbit_2018,tamai_high_resolution_2019,cao_tree_2021}, which is relevant primarily near band crossings. Because such crossings occur away from the Fermi surface in the majority of the Brillouin zone, we neglect SOC in our calculations.  
\\
\indent
%
Our DFT and DFPT calculations are carried out using {\sc Quantum Espresso}~\cite{Giannozzi_Quantum_2009} with 10 \!$\times$\! 10 \!$\times$\! 10 $\mathbf{k}$-point and 5 \!$\times$\! 5 \!$\times$\! 5 $\mathbf{q}$-point grids, and then projected onto Wannier orbitals using the {\sc Wannier90} code~\cite{arash_updated_wannier90_2014}. We use the {\sc Perturbo} code for Wannier interpolation of the electronic structure, phonon modes, and $e$-ph coupling, and for computing the $e$-ph self-energy, spectral functions, and transport properties~\cite{zhou_perturbo_2021}. The transport calculations use a 60 $\times$ 60 $\times$ 60 fine $\mathbf{k}$-point grid and 10$^5$ $\mathbf{q}$-points randomly sampled in the Brillouin zone. Transport results were converged with respect to the $\mathbf{k}$-point grid density, number of $\mathbf{q}$-points sampled, the energy window in which the $\mathbf{k}$-points were chosen, and the frequency grid of the self-energy and spectral functions. 
%
%
\section{Results}
\subsection{Electronic structure and $e$-ph coupling}
\vspace{-10pt}
The electronic structure of SRO is strongly renormalized by electron correlations.  
%
%
\begin{figure}
    \centering
    \includegraphics[width = 0.5 \textwidth]{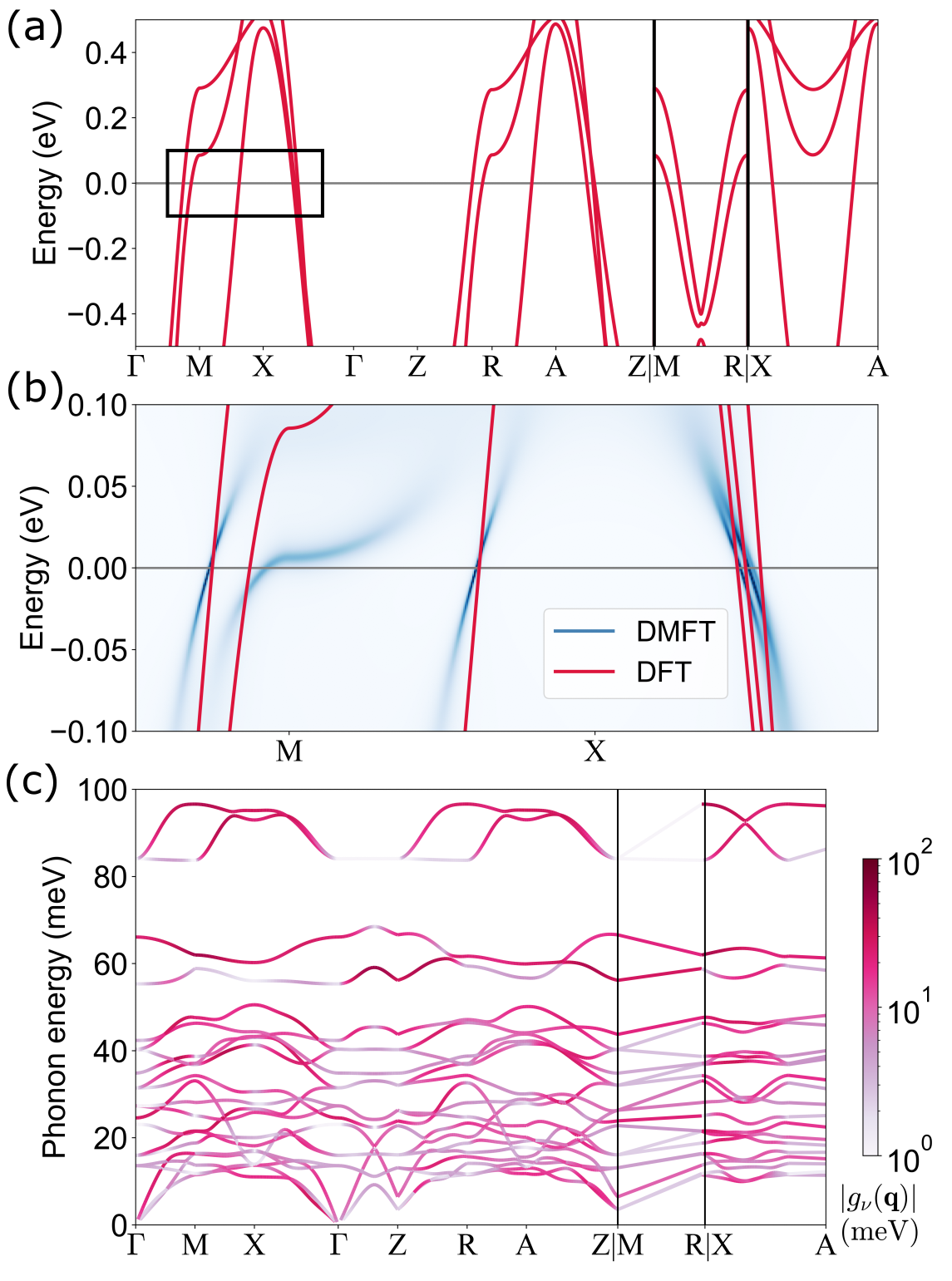}
    \caption{Electronic structure and $e$-ph coupling in SRO. (a) DFT band structure for the $t_{2g}$ $d$ bands included in the Wannierization. (b) DMFT spectral function at 77 K, plotted in an enlargement of the black rectangle in (a), showing significant renormalization relative to the DFT bands near the Fermi energy (shown with a horizonal line). (c) Phonon dispersion colored based on the $e$-ph coupling strength, which is computed as $|g_\nu(\mathbf{q})| = (\sum_{mn}|g_{mn\nu}(\mathbf{k} = M,\mathbf{q})|^2/N_b)^{1/2}$~\cite{zhou_perturbo_2021} by averaging over $N_b=3$ bands.}
    \label{fig:bandsephmat}
\end{figure}
The DFT band structure near the Fermi energy consists of relatively narrow $d$ bands [Fig.~\ref{fig:bandsephmat}(a)]. 
The $e$-$e$ interactions significantly decrease the $d$-band Fermi velocity relative to DFT, as captured by the DMFT electronic spectral functions  [Fig.~\ref{fig:bandsephmat}(b)]. In the 50$-$310~K temperature range studied here, the $e$-$e$ interactions also reduce the spectral weight $Z$ of the electronic states on the Fermi surface $-$ to $Z\approx0.2$ for the $d_{xy}$ and $Z\approx0.3$ for the $d_{xz}$ and $d_{yz}$ bands~\cite{tamai_high_resolution_2019}  
$-$ and cause a large broadening of the associated spectral functions. 
These effects are a signature of strong electron correlations in SRO and highlight the need to treat the electronic structure beyond the band picture. 
\\
\indent
The phonon dispersion with a color map of the $e$-ph coupling strength is shown in Fig.~\ref{fig:bandsephmat}(c). The dispersions are calculated from real space interpolation of DFPT forces~\cite{zhou_perturbo_2021} and are in good agreement with inelastic neutron scattering measurements~\cite{braden_lattice_dynamics_2007}. We find that the $e$-ph interactions are overall relatively weak in SRO, with coupling strength $|g| < 100$~meV for all phonon modes. 
These values are significantly smaller than in insulating transition metal oxides $-$ for example, CoO and SrTiO$_3$~\cite{zhou_predicting_2019,zhou_abinitio_2021} $-$ where the long-range Fr\"{o}hlich interaction with longitudinal optical (LO) modes~\cite{zhou_predicting_2019,zhou_abinitio_2021}  reaches coupling strengths of order $|g| \approx 1$~eV. Due to its metallic character, such polar LO phonons are screened out in SRO, resulting in short-range $e$-ph interactions with weaker coupling strengths.
%
%
\subsection{Electron-phonon self-energy calculations}
\vspace{-10pt}
\label{sec:ephfromspec}
\indent
\begin{figure}[t]
\begin{tikzpicture}
\begin{feynman}
\vertex (a);
\vertex[right=3.0 cm of a] (b) ;
\vertex[right=1.5 cm of b] (c);
\vertex[right=3.0 cm of c] (d) ;
\vertex[left= 0.2 cm of a,above= 1.5 cm of a] (l1) {(a)};
\vertex[left= 0.2 cm of c,above= 1.5 cm of c] (l2) {(b)};
\vertex[above = 0.06 cm of c] (e);
\vertex[above = 0.06 cm of d] (f); 
\diagram*{ 
 (a) -- [fermion, edge label'=\(G^{0}_{m\mathbf{k}+\mathbf{q}}\)] (b),
  (a) -- [photon, half left, edge label=\(D_{\nu q}\)] (b),
  (c) -- [fermion, thick, edge label'=\(G^{\text{DMFT}}_{m\mathbf{k}+\mathbf{q}}\)] (d),
  (e) -- [fermion, thick] (f),
  (c) -- [photon, half left, edge label=\(D_{\nu q}\)] (d)
};
\end{feynman}
\end{tikzpicture}
    \caption{Feynman diagrams for the $e$-ph self-energy $\Sigma^{\textit{e}\text{-ph}}_{n\mathbf{k}}$, computed with (a) the noninteracting electron Green's function and (b) the DMFT electron Green's function. The vertices carry a factor equal to the $e$-ph coupling, $g_{mn\nu}(\mathbf{k},\mathbf{q})$.}
    \label{fig:sefeynmandiagrams}
\end{figure}
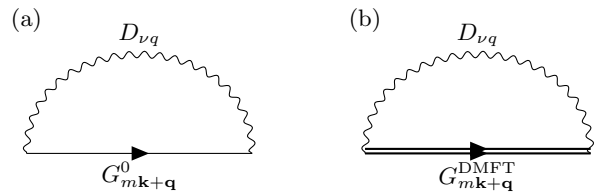
We describe the $e$-$e$ and $e$-ph interactions using the corresponding self-energies. 
When $e$-ph coupling is relatively weak, the $e$-ph interactions are well described by the lowest-order (so called Fan-Migdal) self-energy~\cite{mahanManyParticle2000}
%
\begin{multline}
    \Sigma^{\textit{e}\text{-ph}}_{n\mathbf{k}}(i\omega_n) = \sum_{m\nu\mathbf{q}}\sum_{i\nu_n} |g_{mn\nu}(\mathbf{k},\mathbf{q})|^2  \times \\ G_{m\mathbf{k}+\mathbf{q}}(i\omega_n - i\nu_n) D_{\nu\mathbf{q}}(i\nu_n),
    \label{eq:semigdalmatsubara}
\end{multline}
written as a convolution of electron and phonon propagators, $G_{n\mathbf{k}}(i\omega_n)$ and $D_{\nu\mathbf{q}}(i\nu_n)$ respectively, using Matsubara frequencies $i\omega_n$ for electrons and $i\nu_n$ for phonons; $g_{mn\nu}(\mathbf{k},\mathbf{q})$ are $e$-ph coupling \mbox{matrix elements~\cite{zhou_perturbo_2021}.} 
\\
\indent
In current first-principles calculations, this expression is evaluated using noninteracting electron (and phonon) Green's functions, tacitly assuming that the electron spectral functions consist of a sharp QP peak. In this approximation, the $e$-ph self-energy on the real-frequency axis $\omega$, computed as in Fig.~\ref{fig:sefeynmandiagrams}(a) 
with noninteracting electron Green's functions $G_{m \mathbf{k}+\mathbf{q}}^0(z) = (z-\varepsilon_{m \mathbf{k}+\mathbf{q}})^{-1}$, becomes~\cite{mahanManyParticle2000,bernardi2016first}
\begin{multline}
  \Sigma^{\textit{e}\text{-ph}}_{n\mathbf{k}}(\omega,T) = \sum_{m \nu \mathbf{q}}|g_{mn\nu}(\mathbf{k},\mathbf{q})|^2 
  \times \\ \left[\frac{N_{\nu\mathbf{q}}(T) + f_{m\mathbf{k}+\mathbf{q}}(T)}{\omega - \varepsilon_{m\mathbf{k}+\mathbf{q}} + \omega_{\nu\mathbf{q}} + i\eta} + \frac{N_{\nu\mathbf{q}}(T) + 1 - f_{m\mathbf{k}+\mathbf{q}}(T)}{\omega - \varepsilon_{m\mathbf{k}+\mathbf{q}} - \omega_{\nu\mathbf{q}} + i\eta}\right],
  \label{eq:semigdalband}
\end{multline}
where $\omega_{\nu\mathbf{q}}$ are phonon energies while $N_{\nu\mathbf{q}}(T)$ are equilibrium occupation numbers for phonons and $f_{n\mathbf{k}}(T)$ for electrons at temperature $T$. 
When DFT is used to obtain the band structure $\varepsilon_{n\mathbf{k}}$, we denote this self-energy approximation as $e$-ph@DFT. 
\\
\indent
In systems where DFT fails to describe the electronic structure, several methods are used to correct the QP band dispersion and $e$-ph coupling, including GW~\cite{li_many-body_2020,chang_2022}, DFT+U~\cite{zhou_abinitio_2021}, and DFT with hybrid functionals~\cite{hybrid_galli}. Therefore, as previously shown in the literature, the $e$-ph self-energy in Eq.~(\ref{eq:semigdalband}) can be computed using the \lq\lq best available'' QP band structure and $e$-ph couplings from one of these methods~\cite{jhalani_prl, li_many-body_2020, hybrid_galli, lewandowski_does_2021,chang_2022}. Here, for example, by fitting the QP peaks of the DMFT spectral functions we obtain an improved DMFT band structure and use it to compute the self-energy in Eq.~(\ref{eq:semigdalband}). Below we refer to this level of theory as $e$-ph@DMFT bands. 
\\
\indent
However, even with these improved schemes, the standard $e$-ph self-energy in Eq.~(\ref{eq:semigdalband}) misses key features of the electronic structure in correlated materials. These include the QP  peak broadening, finite QP weight, and any satellite peaks, background, or spectral weight redistribution outside the QP peak, all of which are encoded in the electron spectral function. 
%
Improving the QP band structure or $e$-ph couplings in Eq.~(\ref{eq:semigdalband}) addresses only part of these renormalization effects because it places the full spectral weight on the electronic band states, neglecting spectral weight redistribution from $e$-$e$ interactions. 
The inadequacy of this approximation is apparent in correlated systems like SRO, where the QP weight of the $d$ bands is only $Z\approx$ 0.2$-$0.3.  Going beyond these limitations requires computing the $e$-ph self-energy directly from the electronic spectral function dressed by electron correlations. 
\\ 
\indent
To address this challenge, we develop a method to calculate the $e$-ph self-energy from the DMFT electron Green's function, $G^{\text{DMFT}}_{n\mathbf{k}}$. Starting from the Lehmann representation~\cite{mahanManyParticle2000}, and using a band-diagonal DMFT self-energy, we write:
\begin{equation}
\label{eq:Gdmft}
    G^{\text{DMFT}}_{n\mathbf{k}}(z) = \int\! d\omega\, \frac{A^{\text{DMFT}}_{n\mathbf{k}}(\omega)}{z - \omega}\,,
\end{equation}
where $A^{\text{DMFT}}_{n\mathbf{k}}(\omega) = - \mathrm{Im}(G^{\text{DMFT}}_{n\mathbf{k}}(\omega))/\pi$ is the DMFT spectral function.  
We substitute this expression in Eq.~(\ref{eq:semigdalmatsubara}) and follow the usual derivation of the lowest-order $e$-ph self-energy~\cite{coleman2015introduction}. After analytic continuation to the real frequency axis, $i\omega_n \rightarrow \omega + i\eta$, we obtain:
\begin{multline}
  \Sigma^{\textit{e}\text{-ph}}_{n\mathbf{k}}(\omega,T) = \sum_{m,\nu,\mathbf{q}}|g_{mn\nu}(\mathbf{k},\mathbf{q})|^2 
  \int d\omega' A^{\text{DMFT}}_{m,\mathbf{k}+\mathbf{q}}(\omega') \\ \left[ \frac{(N_{\nu \mathbf{q}}(T) + f(\omega',T))}{\omega - \omega' + \omega_{\nu \mathbf{q}} + i\eta} + \frac{(N_{\nu \mathbf{q}}(T) + 1 - f(\omega',T))}{\omega - \omega' - \omega_{\nu \mathbf{q}} + i\eta} \right].
\label{eq:sigmamigdalspec}
\end{multline}
\indent
This $e$-ph self-energy, written as an integral of the DMFT spectral function, captures on equal footing the key factors mentioned above $-$ band renormalization, finite QP weight and broadening, and background or satellite contributions caused by electron correlations $-$ all of which are included in the DMFT spectral function.
This approximation for the $e$-ph self-energy, here referred to as $e$-ph@$G^\text{DMFT}$, is shown diagrammatically in Fig.~\ref{fig:sefeynmandiagrams}(b). Its numerical evaluation is discussed in Appendix~\ref{app:selfenergycalc}. (Note that in this work we do not renormalize the $e$-ph coupling, $g_{mn\nu}(\mathbf{k},\mathbf{q})$, using DMFT; the role of that renormalization is discussed in Section~\ref{sec:discussion}.)\\ 
%
 
\subsection{Electron-phonon and DMFT self-energies}
\begin{figure*}[ht!]
    \centering
    \includegraphics[width = 1.0\textwidth]{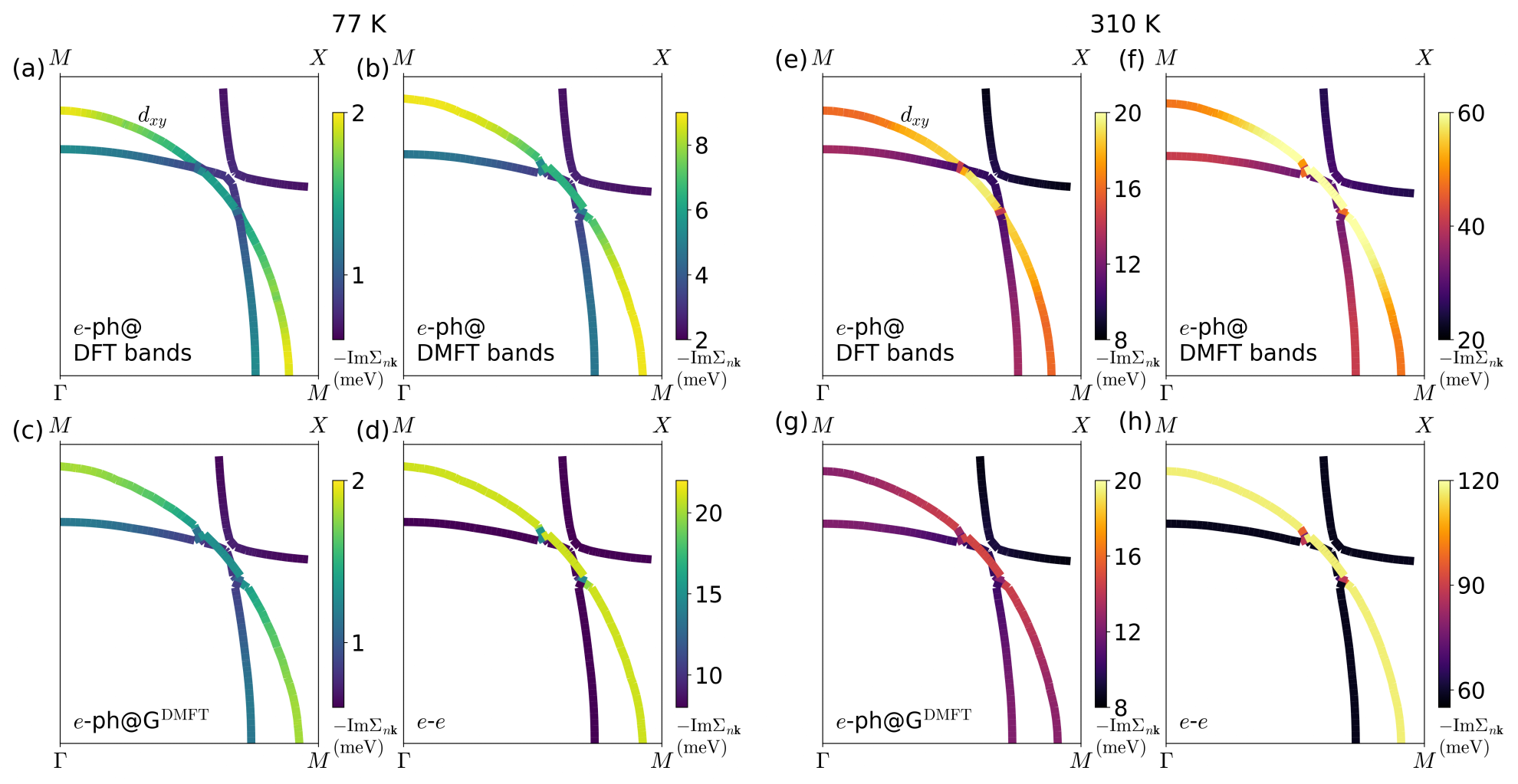}
    \caption{
    Imaginary part of the $e$-ph and $e$-$e$ self-energies in SRO, plotted on cuts of the Fermi surface at 77~K (left) and 310~K (right). We plot the $e$-ph self-energy obtained using (a) the DFT electronic structure ($e$-ph@DFT), (b) the DMFT band dispersion ($e$-ph@DMFT bands), and (c) the DMFT spectral function using Eq.~(\ref{eq:sigmamigdalspec}) ($e$-ph@$G^\text{DMFT}$). \mbox{(d) DMFT} $e$-$e$ self-energy, shown here for comparison. The corresponding self-energies at 310 K are plotted in (e)-(h) respectively. 
    }
    \label{fig:sefsplot}
\end{figure*}

\vspace{-10pt}
We analyze the $e$-ph self-energy obtained with these different approximations and compare it with the DMFT $e$-$e$ self-energy. Figure~\ref{fig:sefsplot} shows these quantities on a cut of the Fermi surface in the $xy$-plane (The real and imaginary frequency dependent self-energy at select $\mathbf{k}$-points is plotted in SM~\cite{supplementary_material}).
%
In each plot, the middle band with stronger $e$-ph and $e$-$e$ coupling has $d_{xy}$ character while the other two bands have $d_{xz}$ and $d_{yz}$ characters. 
The $e$-ph self-energy, which depends on both electronic orbital and momentum $\mathbf{k}$, varies by a factor of $\sim$1.5 on the Fermi surface at 77~K and 310~K. This $\mathbf{k}$-dependence in the self-energy also changes direction as a function of temperature in some parts of the Brillouin zone. For example in the of the d$_{xy}$ band it is greater near the $\Gamma-$M ($\Gamma-$X) axis at low (high) temperature, due to a change of available $e$-ph inter- and intra-band scattering processes.
Our DMFT self-energy also depends on orbital character, but it is local and thus $\mathbf{k}$-independent by construction; it can accurately capture electron correlations in SRO~\cite{tamai_high_resolution_2019}.\\
\indent
Comparing our different $e$-ph self-energy approximations sheds light on the physics they capture. 
In the picture of renormalized QP bands~\cite{coleman2015introduction}, the Fermi velocity decreases by a factor of $1/Z$ (in SRO, $1/Z$ is $\sim$3 for $d_{xz}$ and $d_{yz}$, and $\sim$5 for $d_{xy}$  bands~\cite{tamai_high_resolution_2019}), and the QP weight also decreases by the same factor, leaving the density of states (DOS) at the Fermi energy unchanged. 
This QP picture holds reasonably well in SRO.
The $e$-ph self-energy from DFT bands ($e$-ph@DFT), shown in Fig.~\ref{fig:sefsplot}(a) at 77~K and in Fig.~\ref{fig:sefsplot}(e) at 310~K, misses all these renormalization effects and treats the material as weakly correlated. 
Computing the $e$-ph self-energy in Eq.~(\ref{eq:semigdalband}) with the renormalized DMFT band structure ($e$-ph@DMFT bands), which is obtained by fitting the QP peaks of DMFT spectral functions, 
artificially enhances the $e$-ph self-energy by a factor of 3$-$5, as shown in Fig.~\ref{fig:sefsplot}(b) at 77~K and in Fig.~\ref{fig:sefsplot}(f) at 310~K. This artifact is a consequence of using a band structure with Fermi velocity decreased by a factor of $1/Z \approx 3 - 5$, which increases the DOS at the Fermi energy, and thus also increases $\text{Im}\Sigma^{\rm e-ph}_{n\mathbf{k}}$ in Eq.~(\ref{eq:semigdalband}) by the same factor. 
\\
\indent
However, the DOS at the Fermi energy is nearly unchanged in the renormalized electronic structure because the decrease in Fermi velocity is compensated by a corresponding decrease in QP weight. 
This physics is correctly described by Eq.~(\ref{eq:sigmamigdalspec}), where the imaginary part of the $e$-ph self-energy is roughly proportional to the DOS at Fermi energy written as $D(E_F) \propto \sum_{n\mathbf{k}} \int d\omega' A_{n\mathbf{k}}(\omega') \delta(E_F - \omega')$, which captures changes in the QP weights. 
Therefore, as we show in Fig.~\ref{fig:sefsplot}(c) at 77~K and in Fig.~\ref{fig:sefsplot}(g) at 310~K, the more accurate $e$-ph self-energy computed from the DMFT spectral functions using Eq.~(\ref{eq:sigmamigdalspec}) ($e$-ph@$G^\text{DMFT}$ method) removes the artificial enhancement introduced by using the \lq\lq best-available band structure\rq\rq~approach.  
%
%
\\
\indent
Overall, the $e$-ph self-energy obtained from DFT bands in Fig.~\ref{fig:sefsplot}(a) and DMFT Green's functions in Fig.~\ref{fig:sefsplot}(c) show similar magnitude and trends at 77~K mainly because the DOS at Fermi energy is nearly unchanged in DFT and DMFT. However, at higher temperatures the effects of broadening in the DMFT spectral functions become more pronounced; for example, at 310~K we find a decrease in $\mathbf{k}$-dependence in the $e$-ph self-energy from DFMT Green's function, as shown by comparing Fig.~\ref{fig:sefsplot}(e) and \ref{fig:sefsplot}(g). We attribute this difference to smearing of sharp features in the electronic structure by the spectral width, which is not captured by the DFT-based calculation; this effect is more pronounced in the strongly correlated $d_{xy}$ band. 
Our results show the importance of computing the $e$-ph self-energy from the electron spectral functions in correlated materials to capture the subtle interplay of $e$-ph and $e$-$e$ interactions.\\

\subsection{Spectral functions}
\label{sec:spectral}
\begin{figure}[t!]
    \centering
    \includegraphics[width = 0.5 \textwidth]{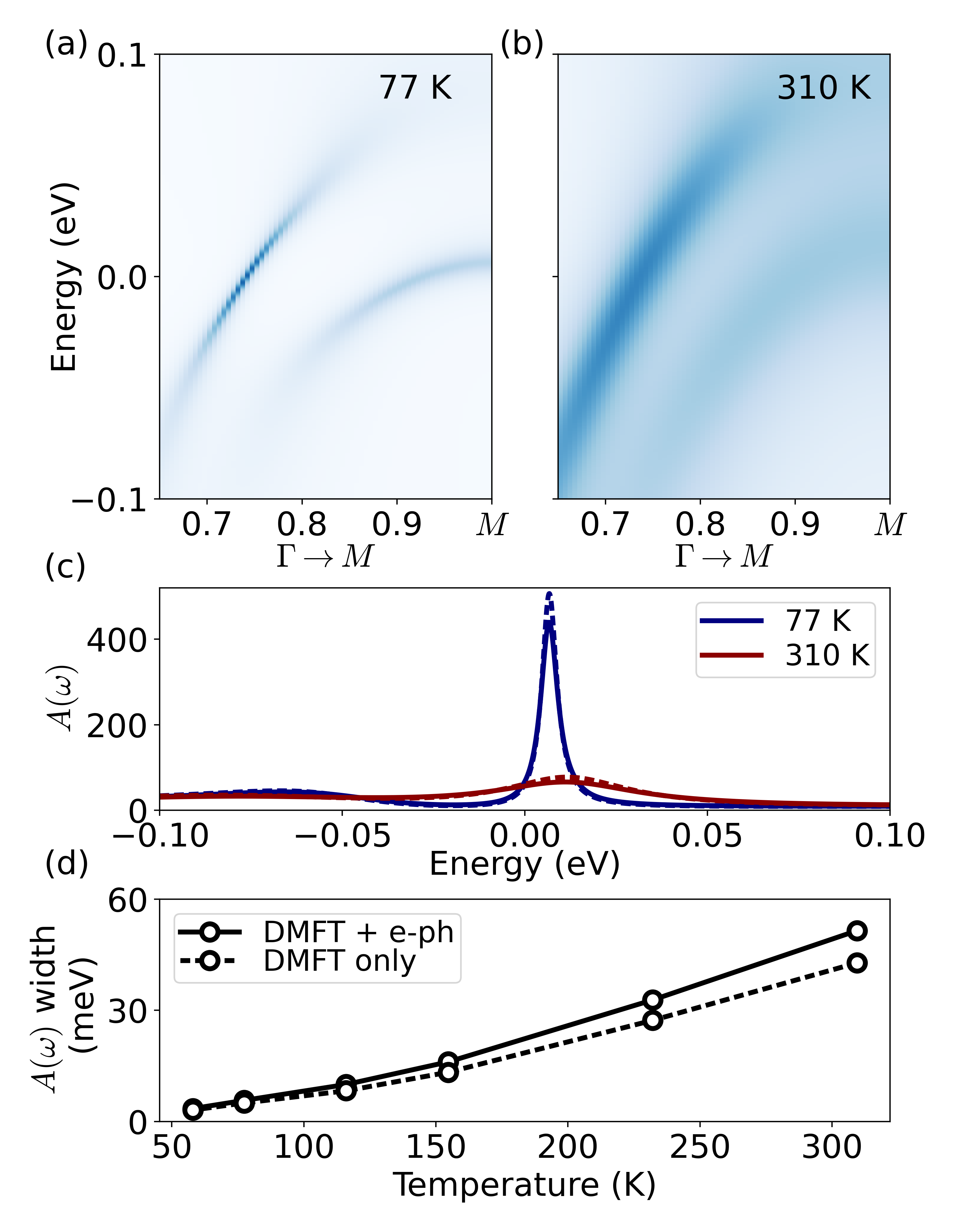}
    \caption{Electronic spectral functions in the $\Gamma \rightarrow M$ direction computed using the DMFT $e$-$e$ plus $e$-ph@$G^{\text{DMFT}}$ self-energies at (a) 77 K and (b) 310 K. (c) Spectral functions at $\mathbf{k} = 0.75$M showing DMFT quasiparticle peaks (dashed lines) broadened by the inclusion of $e$-ph interactions (solid lines) at two temperatures. (d) Calculated full-width at half maximum for the spectral functions in (c), showing quantitatively the small $e$-ph broadening effect.}
    \label{fig:spec}
\end{figure}

\vspace{-10pt}
We obtain the spectral function including both $e$-$e$ and $e$-ph interactions using the Green's function in Eq.~(\ref{eq:Gtotal}):
\vspace{2pt}
\begin{equation}
    A_{n\mathbf{k}}(\omega,T) = -(1/\pi)\,\text{Im}\,G_{n\mathbf{k}}(\omega,T). 
    \label{eq:spectral}\vspace{2pt}
\end{equation}
%
To evaluate this expression, our most accurate approximation consists of first computing the DMFT $e$-$e$ self-energy starting from DFT and then obtaining the $e$-ph self-energy from the DMFT spectral function. This \mbox{$e$-ph@$G^{\rm DMFT}$} approach is justified by the adiabatic approximation: the fast $e$-$e$ interactions renormalize the electronic states, and the slow nuclear motions governing $e$-ph coupling occur in this renormalized ground state.
%
%
\\
\indent 
Figures~\ref{fig:spec}(a)$-$(b) show the electron spectral functions computed in SRO with this approach 
 and plotted in the $\Gamma\! \rightarrow\! M$ direction at 77~K and 310~K. The main features are consistent with photoemission and transport measurements~\cite{tamai_high_resolution_2019, mackenzie_quantum_1996, bergemann_detailed_2000} $-$ in particular, the spectral functions remain fairly sharp near the Fermi energy below $\sim$100~K, indicating well-defined QP excitations at low temperature~\cite{deng_shining_2014}. 
 The spectral broadening increases at higher temperatures, further demonstrating the importance of computing $e$-ph interactions beyond the band picture. Figures~\ref{fig:spec}(c)$-$(d) compare the spectral functions and their QP peak broadening at a fixed $\mathbf{k}$-point ($\mathbf{k} \!=\! 0.75 \text{M}$) with and without the inclusion of $e$-ph interactions. 
We find that the $e$-ph interactions broaden the QP peaks only slightly, consistent with the overall weak $e$-ph coupling in SRO. Increasing the temperature, rather than including $e$-ph interactions, is the main factor responsible for broadening. 
We find similar trends when analyzing spectral functions in the $\Gamma \!\rightarrow\! X$ direction~\cite{supplementary_material}. These results point to a dominant role of $e$-$e$ interactions in SRO. 
%
%
\\
\indent
%



\subsection{Transport}
\label{sec:transport}
\vspace{-10pt}
We study electrical transport in SRO in the characteristic bad metallic regime~\cite{Georges_strong_correlations_2013}, focusing on understanding the roles of $e$-$e$ and $e$-ph interactions. 
We compute the optical conductivity from our combined DMFT plus $e$-ph spectral function in Eq.~(\ref{eq:spectral}), using Green-Kubo theory without current-vertex corrections~\cite{mahanManyParticle2000}:
\begin{multline}
    \sigma_{\alpha\beta}(\omega,T) = \frac{\pi \hbar e^2}{V_{uc}} \int d\omega' \frac{f(\omega',T) - f(\omega'+\omega,T)}{\omega} 
    \times\\ \sum_{nk}v^\alpha_{n\mathbf{k}}v^\beta_{n\mathbf{k}}A_{n\mathbf{k}}(\omega',T)A_{n\mathbf{k}}(\omega' + \omega,T)
\label{eq:greenkubo}
\end{multline}
where $v_{n\mathbf{k}}$ are band velocities, $\alpha$ and $\beta$ are Cartesian directions, and $V_{\rm uc}$ is the unit cell volume; the dc resistivity tensor is obtained as $\rho_{\rm dc}(T) = \sigma^{-1}(\omega \to 0,T)$. 
%
%
\begin{figure*}[!ht]
    \centering
    \includegraphics[width = 0.95\textwidth]{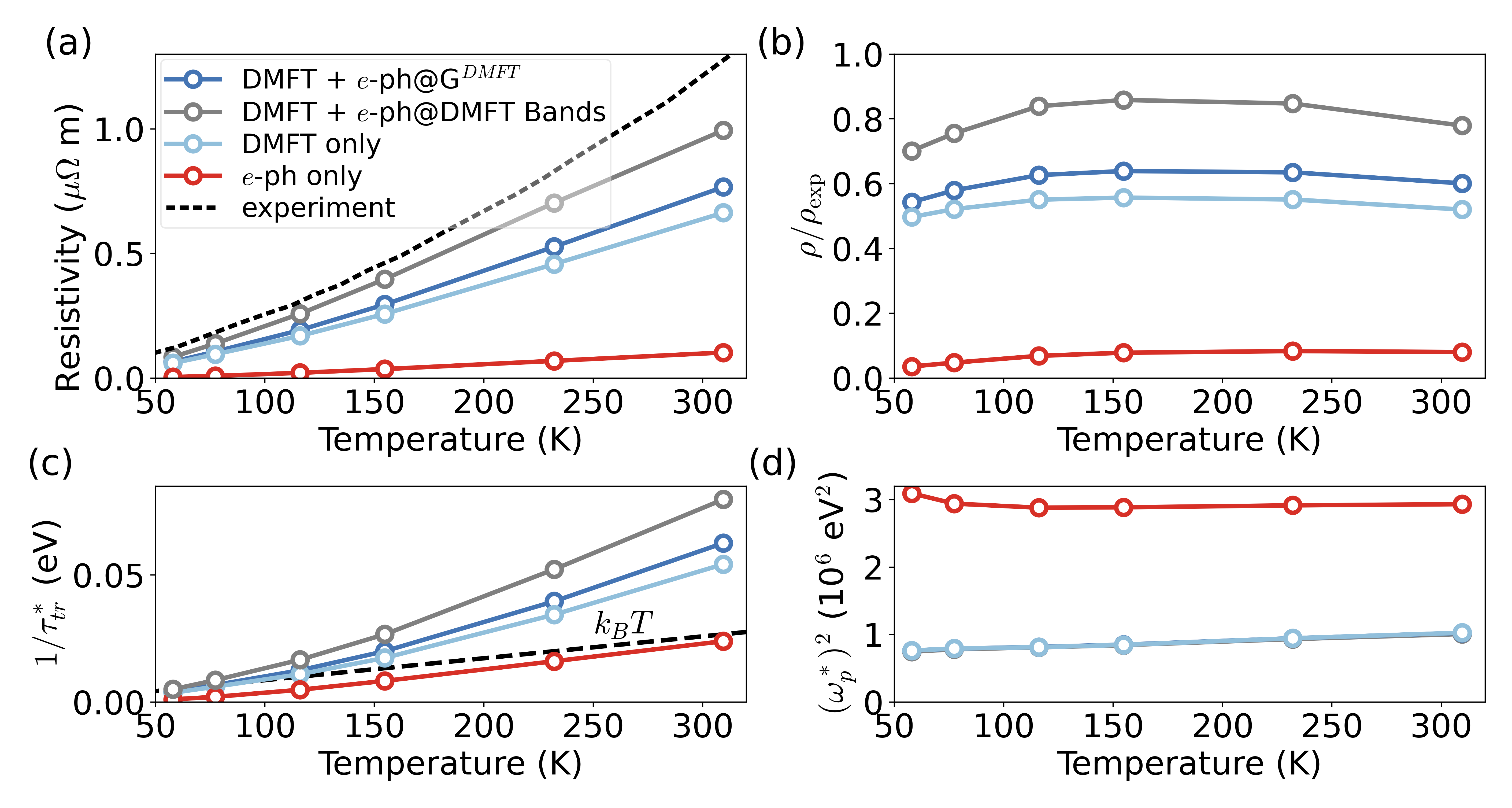}
    \caption{Transport properties as a function of temperature in SRO. (a) In-plane resistivity, comparing DMFT+$e$-ph, DMFT-only, and $e$-ph-only calculations to experimental data from Ref.~\cite{tyler_high-temperature_1998}. (b) Same as (a), but plotted as a fraction of the experimental resistivity. (c) Effective scattering rate, and (d) effective plasma frequency. In the legend, DMFT only and e-ph only refer to using, respectively, only the DMFT or $e$-ph self-energies in the Green's function, spectral function, and Green-Kubo formula.}
    \label{fig:resopcond} 
\end{figure*}
This approximation has been used (separately) to study $e$-ph limited transport in oxides and organic crystals~\cite{zhou_predicting_2019, chang_2022} and $e$-$e$ limited transport in correlated  metals~\cite{deng_transport_2016,haule_dynamical_2010,mravlje_thermopower_2016,pourovskii_electron_electron_2017}; different from these previous studies, here we focus on combining the $e$-$e$ and $e$-ph interactions and studying their interplay.
\\
\indent 
Figure~\ref{fig:resopcond}(a) shows the resistivity in the $xy$-plane as a function of temperature. We compare the resistivity limited by $e$-$e$ interactions alone, computed from the Green's function including only the DMFT self-energy, with  calculations including both $e$-$e$ and $e$-ph interactions in the Green's function. When both interactions are included, we compare results from the more accurate $e$-ph@$G^{\text{DMFT}}$ approach with calculations using Eq.~(\ref{eq:semigdalband}) with the best-available band structure ($e$-ph@DMFT bands method). Experimental values are also shown for comparison~\cite{tyler_high-temperature_1998}. For convenience, Fig.~\ref{fig:resopcond}(b) shows the same results expressed as a fraction of the experimental resistivity.  
%
\\
\indent
The calculation including only DMFT $e$-$e$ interactions correctly predicts the order of magnitude of the experimental resistivity, providing a resistivity smaller than experiment by a factor of 2 in the entire temperature range studied here (50$-$310~K). This calculation is comparable with results from Deng~\textit{et al.}~\cite{deng_transport_2016}, who used a different scheme to construct the correlated subspace in the DMFT calculation and obtained somewhat larger values for the $e$-$e$ limited resistivity due to stronger interaction parameters $U$ and $J$. 
In contrast, the resistivity limited by the $e$-ph interactions alone (computed with the $e$-ph@$G^{\text{DMFT}}$ approach) is only $\sim$10\% of the \mbox{experimental} value.
\\
\indent
When combining $e$-$e$ and $e$-ph interactions through the $e$-ph@$G^{\text{DMFT}}$ approach, the resulting resistivity is the sum of the individual $e$-$e$ and $e$-ph limited resistivities, a behavior known as Matthiessen's rule~\cite{Ziman}. Therefore, adding $e$-ph interactions improves the agreement with experiment and predicts a resistivity equal to $\sim$60\% of the experimental value between 50$-$310~K. 
In contrast, using only an improved QP band structure ($e$-ph@DMFT bands) artificially enhances the $e$-ph interactions, leading to an incorrect prediction that the $e$-ph contribution is 30\% of the experimental resistivity [see Fig.~\ref{fig:resopcond}(b)]. 
Note that our most accurate calculation still underestimates the experimental resistivity by $\sim$40\%. The possible origin of this \lq\lq missing resistivity\rq\rq~ is discussed below. 
\\
\indent
To better understand the contributions to the resistivity, we analyze our transport results using an effective Drude model~\cite{deng_shining_2014}, writing the resistivity as $\rho_{\rm dc} = 4\pi / [\tau_{\rm tr}^*(\omega_{\rm p}^*)^2]$, where $1/\tau_{\rm tr}^*$ is an effective scattering rate and $\omega_{\rm p}^*$ is an effective plasma frequency. 
We extract these quantities from our computed optical conductivity $\sigma(\omega)$ using~\cite{deng_shining_2014} 

\begin{equation}
    \tau_{\rm tr}^* = -\frac{2}{\pi \sigma_{\rm dc}}\int_0^\infty \frac{1}{\omega}\frac{\partial \mathrm{Re}(\sigma(\omega))}{\partial \omega} d\omega
\end{equation}
and 
\begin{equation}
    (\omega_{\rm p}^*)^2 = \frac{-2\pi^2\sigma_{\rm dc}^2}{ \int_0^\infty \frac{1}{\omega}\frac{\partial \mathrm{Re}(\sigma(\omega))}{\partial \omega} d\omega}.
\end{equation}
\\
\indent
Figure~\ref{fig:resopcond}(c) shows the effective scattering rate $1/\tau_{tr}^*$ and Fig.~\ref{fig:resopcond}(d) the plasma frequency $(\omega_p^*)^2$ obtained from this analysis.  
These results confirm that $e$-$e$ interactions dominate electron scattering and lead to a decreased Fermi velocity (and thus plasma frequency) by renormalizing the band structure. 
%
The effective $e$-ph scattering rate is significantly smaller (and the plasma frequency greater) than for $e$-$e$ interactions, consistent with the small $e$-ph contribution to the resistivity and negligible band renormalization (relative to DFT) from $e$-ph interactions. 
In addition, most of the temperature dependence of the resistivity comes from the effective scattering rates,  while the effective plasma frequency depends weakly on temperature.
%
%
%
%
\\
\indent
 Figure~\ref{fig:resopcond}(c) additionally compares the effective scattering rates with the Planckian limit $\Gamma_{\mathrm{p}} = k_{\rm{B}}T$~\cite{hartnoll_theory_2015}. Interestingly, the $e$-ph scattering rate is lower than $\Gamma_{\mathrm{p}}$ and approaches it near 310 K, whereas the scattering from $e$-$e$ interactions exceed $\Gamma_{\mathrm{p}}$ above $\sim$100~K, reaching $\sim2 k_{\rm B}T$ at 310~K.
%
The weaker $e$-ph scattering in SRO contrasts the cases of insulating oxides such as SrTiO$_3$ and CoO~\cite{zhou_predicting_2019,zhou_abinitio_2021}, where pronounced polaron effects are present $-$ both materials exhibit polaron satellites, and in SrTiO$_3$ the effective $e$-ph scattering rate exceeds the Planckian limit~\cite{zhou_predicting_2019}.\\

%
%
\section{Discussion}
\label{sec:discussion}
\vspace{-10pt}
Our most accurate transport calculation underestimates the in-plane resistivity in SRO by $\sim$40\% relative to experiments. The origin of this discrepancy deserves a detailed discussion. 
First, we verify that our lowest-order (Fan-Migdal) treatment of $e$-ph interactions is sufficient in SRO by recomputing the resistivity with an $e$-ph cumulant method that can describe higher-order $e$-ph interactions and transport in the presence of polarons~\cite{zhou_predicting_2019, chang_2022}. 
Consistent with the effective screening of polar phonons in metals, we find that the $e$-ph limited transport is nearly identical in the lowest-order and cumulant calculations~\cite{supplementary_material}.
\\
\indent
%
Second, our Green-Kubo calculations neglect current-vertex corrections, which typically play a small role in $e$-ph limited transport in metals. While vertex corrections are difficult to compute in Green-Kubo theory~\cite{park_prl,park_prb}, they can be quantified in the semiclassical limit using the Boltzmann transport equation (BTE). In particular, including vertex corrections corresponds to a full solution of the BTE, obtained here with an iterative approach (ITA)~\cite{zhou_perturbo_2021}, while neglecting vertex corrections is equivalent to the relaxation time approximation (RTA), to which the Green-Kubo formula reduces in the weak-coupling limit~\cite{zhou_predicting_2019}.
Therefore, we can estimate the effect of vertex corrections by comparing our $e$-ph limited resistivity computed with Green-Kubo to BTE calculations using the ITA and RTA, which respectively include and neglect vertex corrections~\cite{park_prb}. From low temperature up to 310~K, the ITA resistivity is $\sim$20\% greater than the RTA value, while the RTA and Green-Kubo calculations agree to within 2\% at 310 K~\cite{supplementary_material}. 
These results indicate that vertex corrections have a modest effect on $e$-ph limited transport. 
\\
\indent
Third, most DMFT studies of transport neglect  vertex corrections from $e$-$e$ interactions, which is strictly correct only in the limit of infinite dimensions~\cite{haule_dynamical_2010}. However, their role is difficult to quantify and often non-negligible~\cite{vucicevic_conductivity_2019}. Fourth, although our DMFT calculations produce band dispersions in excellent agreement with photoemission data~\cite{tamai_high_resolution_2019}, DMFT results can in some cases be sensitive to the method used to construct the local site~\cite{karp_dependence_2021}, and  neglecting nonlocal interactions may remove relevant $e$-$e$ scattering mechanisms which affect the calculated spectral width. Studying these effects is an active area of DMFT research and is beyond the scope of this work. 
\\
\indent
Finally, recent work has shown that DFPT based on semilocal functionals can lead to underestimated $e$-ph coupling in correlated materials~\cite{yin_correlation-enhanced-2013,li_electron_2019}. Yet much of that work has focused on materials with nonlocal correlations, so it is difficult to estimate the size of the effect in SRO,  where correlations are dominantly local~\cite{tamai_high_resolution_2019}. 
It is possible that the $e$-ph coupling and its role in transport are somewhat stronger than captured by our DFPT calculations. 
However, an enhancement of the $e$-ph coupling consistent with those reported to date would not change our main conclusion that the $e$-ph contribution to the resistivity is relatively small $-$ for example, assuming the limit case of a doubling of $e$-ph coupling strength $|g|^2$ would still give only a $\sim$20\% $e$-ph resistivity contribution.
\\
\indent
A qualitative comparison between $e$-$e$ and $e$-ph interactions is also interesting. The $\mathbf{k}$-dependence of the $e$-ph interactions clearly contrasts the local ($\mathbf{k}$-independent) $e$-$e$ interactions. Extending this comparison to other correlated materials may highlight key differences between materials with dominant $e$-$e$ or $e$-ph interactions. We also observe that our $e$-ph@$G^{\mathrm{DMFT}}$ self-energy generally has a weaker $\mathbf{k}$-dependence than predicted by the $e$-ph@DFT method, suggesting that electron correlations suppress nonlocal $e$-ph interactions. This effect may have interesting implications for materials with strong or long-range $e$-ph interactions, including cuprates and correlated insulators.
\\ 
\indent
Lastly, our development of the $e$-ph@$G^{\text{DMFT}}$ method fits within efforts in computational many-body physics to obtain accurate components of Feynman diagrams. For $e$-ph interactions, this includes improving the accuracy of the electron propagator $G$, phonon propagator $D$, and $e$-ph coupling $g$. When viewed this way, there is a clear analogy with the use of different levels of theory in the GW method, where the accuracy can be improved by using the \lq\lq best G\rq\rq~and \lq\lq best W\rq\rq. 
In $e$-ph calculations, methods for Hubbard-~\cite{zhou_abinitio_2021} or DMFT-corrected phonons~\cite{kocer_efficient_lattice_dynamics_2020,khanal_correlation_2020}, as well as anharmonic lattice dynamics~\cite{hellman_lattice_2011,zhou_predicting_2019}, address the phonon propagator $D$, while Hubbard-corrected DFPT~\cite{zhou_abinitio_2021}, GW perturbation theory (GWPT)~\cite{li_electron_2019}, and DFT+DMFT deformation potentials~\cite{mandal_strong_2014} aim to improve the $e$-ph coupling $g$. These effects can be treated at a suitable level of theory for each material; for example, if the lattice is strongly anharmonic, the phonon propagator can be computed using renormalized phonon frequencies, or one can use a full frequency dependent phonon propagator analogous to our use of the DMFT Green's function.
\\
\indent
Our work improves the accuracy of the electron propagator $G$ entering the $e$-ph self-energy diagram, by dressing it with the DMFT self-energy to capture dynamical electron correlations. This provides a scheme where the $e$-ph self-energy is computed from the DMFT spectral functions. Our results highlight the importance of using the spectral functions to compute $e$-ph interactions in correlated materials, and cautions against using just the best available QP band structure $-$ if the dispersion is renormalized but the QP weight is not adjusted, the $e$-ph self-energy becomes inaccurate as we have shown for SRO.

\section{Conclusion}
\vspace{-10pt}
We demonstrated a method to combine strong dynamical correlations described by DMFT with first-principles $e$-ph calculations. This advance enables a quantitative treatment of both $e$-$e$ and $e$-ph interactions and their effect on transport in correlated materials. Our approach computes $e$-ph interactions from the DMFT Green's function, which captures band structure and spectral weight renormalization from strong electron correlations. In SRO, where $e$-ph interactions are relatively weak, we have shown that transport is governed by $e$-$e$ interactions, with $e$-ph interactions contributing only $\sim$10\% of the experimental resistivity in a wide temperature range. 
\\
\indent 
Our work expands the reach of first principles $e$-ph calculations to strongly correlated materials by treating the electronic structure at the level of the spectral function rather than non-interacting bands. Note that our calculations can also use more accurate versions of the phonon propagator (e.g., Hubbard-corrected DFT, DMFT, or anharmonic lattice dynamics) and $e$-ph coupling (e.g., from GWPT or DFPT+U) as an input for calculating interactions, and thus are complementary to these approaches. 
The method presented in this work enables studies of a range of correlated materials, particularly when combined with emerging approaches to treat $e$-ph interactions in materials with spin-orbit coupling~\cite{zhou_perturbo_2021}, magnetism, and Mott insulating gaps~\cite{zhou_abinitio_2021}.  Considering the wide variety of physical parameters in correlated materials $-$ number of relevant orbitals, conducting or insulating, type of magnetic ordering, strength of spin-orbit coupling, etc.~$-$ one expects that the interplay of $e$-$e$ and $e$-ph interactions will vary greatly in these systems, opening wide-ranging possibilities for future work. 

%

\section*{ACKNOWLEDGMENTS}
\vspace{-10pt}
The authors thank Manuel Zingl for sharing DMFT data and for fruitful discussions. 
This work was primarily supported by the National Science Foundation under Grant No. DMR-1750613, which provided for method development, and Grant No. OAC-2209262, which provided for code development. 
D.J.A. and M.B. were partially supported by the AFOSR and Clarkson Aerospace under Grant No. FA95502110460. 
J.-J.Z. acknowledges support from the National Natural Science Foundation of China (Grant No. 12104039). J.M. is supported by the Slovenian Research Agency (ARRS) under Grants No. P1-0044 and J1-2458. This research used resources of the National Energy Research Scientific Computing Center (NERSC), a U.S. Department of Energy Office of Science User Facility located at Lawrence Berkeley National Laboratory, operated under Contract No. DE-AC02-05CH11231.
\newpage
\appendix
\begin{section}{Numerical Evaluation of the Electron-Phonon Self-Energy}
\label{app:selfenergycalc}
\end{section}
\vspace{-10pt}
Computing the $e$-ph self-energy from DMFT spectral functions using Eq.~(\ref{eq:sigmamigdalspec}) requires an additional frequency integral compared to the standard $e$-ph self-energy in Eq.~(\ref{eq:semigdalband}). 
We develop an efficient approach to compute this $e$-ph@$G^{\rm DMFT}$ self-energy. Using the relation
\begin{equation}
    \frac{1}{\omega + i\eta} = P\left(\frac{1}{\omega}\right) -i\pi\delta(\omega),
\end{equation}
where $P$ denotes the principal part, allows us to write the imaginary part of the $e$-ph self-energy as 
\begin{multline}
    \text{Im}\Sigma^{\rm e-ph}_{n\mathbf{k}}(\omega) = -\pi \sum_{m\nu\mathbf{q}} |g_{mn\nu}(\mathbf{k},\mathbf{q})|^2 \\ [ (N_{\nu \mathbf{q}} + f(\omega + \omega_{\nu \mathbf{q}}))A_{m\mathbf{k}+\mathbf{q}}(\omega + \omega_{\nu \mathbf{q}}) \\ + (N_{\nu \mathbf{q}} + 1 - f(\omega - \omega_{\nu \mathbf{q}}))A_{m\mathbf{k}+\mathbf{q}}(\omega - \omega_{\nu \mathbf{q}}) ].
    \label{eq:imsigmaspec}
\end{multline}
This way, the calculation has a computational cost similar to the standard $e$-ph self-energy in Eq.~(\ref{eq:semigdalband}), but it additionally requires the shifted DMFT spectral functions 
\begin{equation}
    A^{\rm DMFT}_{n\mathbf{k}}(\omega' = \omega \pm \omega_{\nu \mathbf{q}}) = -\frac{1}{\pi}\text{Im}\left(\frac{1}{\omega' - \varepsilon_{n\mathbf{k}} - \Sigma^{e-e}_{n\mathbf{k}}(\omega')}\right).
    \label{eq:shiftedspec}
\end{equation}
In DMFT, $\Sigma^{\rm e-e}_{n\mathbf{k}}(\omega)$ is local in the Wannier basis and thus it can be rapidly computed using Eq.~(\ref{eq:sigmaband}) and interpolated to the shifted energies $\omega' = \omega \pm \omega_{\nu \mathbf{q}}$. The real part of the $e$-ph@$G^{\rm DMFT}$ self-energy, which is not computed in this work, can be obtained from the imaginary part using the Kramers-Kronig relations without repeating the sum over phonon modes and momenta.
%
\\
\indent
In our calculations, we obtain the $e$-ph@$G^{\rm DMFT}$ self-energy using a fine energy grid with 0.5 meV spacing in a window of $\pm100$~meV around the Fermi energy ($E_F$), together with a coarser grid with 20 meV spacing in the $E_F\pm$2~eV energy window. We obtain the spectral functions in Eq.~(\ref{eq:shiftedspec}) using Wannier-interpolated band energies and the DMFT self-energy interpolated in frequency from a grid with 4 meV spacing. 
\bibliographystyle{apsrev4-2}
\bibliography{apssamp} 
\end{document}